\documentclass[aps,reprint,prl,superscriptaddress]{revtex4-1}

\usepackage{graphicx}  
\usepackage{bm}        
\usepackage{amssymb,amsmath}
\usepackage{xcolor}
\usepackage{mathtools}
\usepackage{bbm}

\newcommand{\p}{\partial} 
\newcommand{\E}{\mathbb E} 
\newcommand{\erfc}{\operatorname{erfc}}

\newcommand{\R}{\mathbb R}
\newcommand{\C}{\mathbb C}
\newcommand{\prob}{\mathbb P}

\newcommand{\diag}{\operatorname{diag}} 
\renewcommand{\vec}{\operatorname{vec}} 
\newcommand{\tr}{\operatorname{Tr}} 

\renewcommand{\phi}{\varphi} 
\renewcommand{\epsilon}{\varepsilon} 

\DeclarePairedDelimiter{\abs}{\lvert}{\rvert} 

\begin{document}

\raggedbottom

\title{Nonlinearity-generated Resilience in Large Complex Systems}

\author{S. B. Fedeli}\email{sirio.belga\_fedeli@kcl.ac.uk}
\affiliation{Department of Mathematics, King’s College London, London WC2R 2LS, United Kingdom}

\author{Y. V. Fyodorov}\email{yan.fyodorov@kcl.ac.uk}
\affiliation{Department of Mathematics, King’s College London, London WC2R 2LS, United Kingdom}
\affiliation{L.D.Landau Institute for Theoretical Physics, Semenova 1a, 142432 Chernogolovka, Russia}

\author{J. R. Ipsen}\email{jesper.ipsen@unimelb.edu.au}
\affiliation{ARC Centre of Excellence for Mathematical and Statistical Frontiers, School of Mathematics and Statistics, The University of Melbourne, 3010 Parkville, VIC, Australia}

\begin{abstract}
\noindent

 We consider a generic nonlinear extension of May's 1972 model by including all higher-order terms in the expansion around the chosen fixed point (placed at the origin) with random Gaussian coefficients. The ensuing analysis 
 reveals that as long as the origin remains stable, it is surrounded by a ``resilience gap": there are no other fixed points within a radius $r_*>0$ and the system is therefore expected to be resilient to a typical initial displacement small in comparison to $r_*$. The radius $r_*$ is shown to vanish  at the same threshold where the origin loses local stability, revealing a mechanism by which systems close to the tipping point become less resilient. We also find that beyond the resilience radius the number of fixed points in a ball surrounding the original point of equilibrium grows exponentially with $N$, making systems dynamics highly sensitive to far enough displacements from the origin.

\end{abstract}

\date{\today}

\maketitle


The dynamics of large complex systems is often modelled as a nonlinear system of coupled first-order differential equations. By virtue of the Hartman--Grobman Theorem the local stability of a `generic' fixed point of a dynamical system (also known as a `point of equilibrium') may be studied using a first-order (linear) approximation in the vicinity of the fixed point. With this in mind, the highly influential paper~\cite{May1972} by Robert May suggested to study stability of ecosystems with many interacting species by considering the linear system
\begin{equation}\label{may}
\frac {d\bm x}{dt}=-\mu \bm x+\frac\sigma{\sqrt N}\bm \varXi\bm x,
\end{equation}
where $\bm x=(x_1,\ldots,x_N)^T$ is an $N$-dimensional vector representing the state of the system, $\mu$ and $\sigma$ are positive constants, and $\bm \varXi=(\xi_{nm})_{n,m}$ is an $N\times N$ connectivity matrix whose entries $\xi_{nm}$ are i.i.d. random variables with zero mean and unit variance. 
The first term on the right-hand side in~\eqref{may} provides a stability feedback mechanism such that in the absence of interactions ($\sigma=0$) the system relaxes to the origin with the decay rate $\mu$. 
The second term gives (random) pairwise interactions between components, such that components $n$ and $m$ have a mutualistic (competitive) relationship if $\xi_{nm}$ and $\xi_{mn}$ are both positive (negative), while they have a parasitic relationship if $\xi_{nm}$ and $\xi_{mn}$ have opposite signs. The parameter $\sigma/\sqrt N>0$ provides an average interaction strength, with the chosen normalization in $N$ ensuring comparability between the first and second terms in (\ref{may}) as $N\to \infty$. The linear system~\eqref{may} is stable if the real part of all eigenvalues of the connectivity matrix $\sigma\bm \varXi/\sqrt N$ is less than $\mu$, while the system becomes unstable if at least one of the eigenvalues has real part larger than $\mu$.

Performing the ensuing analysis, May found that a generic randomly assembled linear complex system 
for large $N\gg 1$  
is stable with probability one (almost surely) if $\mu>\sigma$ and almost surely unstable if $\mu<\sigma$. This observation is based on insights from random matrix theory. The so-called circular law states that for a matrix $\bm \varXi$ with i.i.d. random entries (mean zero, unit variance, finite fourth moment) the empirical spectral density of $\bm\varXi/\sqrt N$ converges to the uniform distribution in the centred unit disk in the complex plane, and the spectral radius converges to $1$ almost surely. It also implies that the eigenvalue of $\bm \varXi/\sqrt N$ with the maximal real part converges to $1$ almost surely. In full generality this result has only been mathematically rigorously established very recently, see ~\cite{BC2012} and references within, but for the Gaussian case was already largely understood when May published his paper. 
May's paper~\cite{May1972} and subsequent book~\cite{MayBook} sparked a long-lived diversity-complexity debate in ecological community which still not fully settled, see~\cite{AT2015,LMBHD2018} for recent reviews. The ideas behind May's model analysis are not restricted to complex ecosystems, and since then has been applied to a much wider class of large complex systems, such as e.g. stability of large economies, see \cite{Bouchaud2020} and references therein.

There are quite a few obvious limitations in May's original model which allowed many researchers to question implications of his analysis to real world systems.   From the point of view of ecology, May's initial model disregarded food-web structure, such as trophic levels \cite{troph}, modularity \cite{modul}, the feasibility of the chosen equilibrium \cite{feas}, as well as many more subtle but relevant effects, see e.g. \cite{Galla2020}.
Some of the restrictions (e.g. an unrealistic feature of being completely randomly assembled)
can, at least partly, be addressed by introducing an additional network architecture to the model; some account
of this activity can be found in the reviews see \cite{AT2015,LMBHD2018} and references within; for a recent interesting works in this direction see \cite{transient} and \cite{specnetwstab}.
Most of these developments are still possible without going beyond the linear approximation.

  Another layer of criticism addresses the fact that though a model like~\eqref{may} might be sufficient for a crude understanding of local stability of a fixed point in a large complex system, retaining linearity  prohibits any deeper questions about its dynamical behaviour. 
  Most obviously, it is meaningless to ask what happens with the system after the chosen fixed point becomes locally unstable (in the ecological context such loss of stability is frequently called a `tipping point'). Moreover, it is well-known that even  before crossing a `tipping point' complex systems typically become increasingly vulnerable to displacements away from the origin due to loss of an “ecological resilience” (e.g. due to shrinkage in the size of the basin of attraction),  see \cite{catast,DKG12} and references therein. Keeping only the linear approximation prohibits meaningfully addressing any natural (and ecologically relevant) questions about mechanisms behind the system's resilience, including the existence of various scenarios of response to an initial displacement away from the locally stable point of equilibrium.

One of the most natural ways in going beyond the May's model (\ref{may}) is to replace it with a system of $N$ coupled autonomous nonlinear ordinary differential equations (ODE's) given by
\begin{equation}\label{model}
\frac {d\bm x}{dt}=-\mu \bm x+{\bf f}(\bm x)
\end{equation}
where $\bm f=(f_1,\ldots,f_N)^T$ is a random vector field. Such model is of course
extremely general and to proceed with its analysis in a meaningful and controllable way one needs to specify the statistical
properties of the random field $\bm f$. Many interesting models could be mentioned in this context, see e.g.~\cite{Iaroslav2015,BKK2016,GBMA2017,FM2014,HMS2016,Bunin2017,BBC18,Galla2018,RBBC,SG2020}. As an example \cite{WT} provide an attempt to perform May's type of analysis for a special choice of the system (\ref{model}) in the framework of neural network dynamics.
The framework dictated choosing $f_i=\sum_{j}J_{ij}S(x_j)$, with $S(x)$ being an odd sigmoid function representing the synaptic nonlinearity, and $J_{ij}$ taken as i.i.d. centred Gaussian variables representing a synaptic connectivity between neurons $i$ and $j$. Although the ensuing dynamical  system is not easily amenable to a fully controllable analysis, a shrewd semi-heuristic insights revealed the existence of critical coupling threshold beyond which there is an exponential in $N$ growth in the total number of equilibria in such a system, and estimated the rate of that growth.

One of the most advanced and systematic attempts in understanding (\ref{model}) beyond linearity has been
undertaken recently in \cite{FK2016} and its sequel \cite{BFK2020}. The authors exploited an idea of
decomposing the interaction field $\mathbf{f}(\mathbf{x})$ into the sum of longitudinal (curl-free, or gradient) and transversal (divergence free) components. Such a construction generalized the model of a pure gradient-descent relaxation of a particle given by 
$ d\mathbf{x}/dt=-\nabla L$, with $L(\mathbf{x})=\mu{|{\bf x}|^2}/{2}+V({\bf x})$ being the associated Lyapunov function describing the effective relaxation landscape. Choosing $V({\bf x})$  as a homogeneous isotropic Gaussian field in $\mathbb{R}^N$ with prescribed covariance one then gets a problem intimately related with the theory of mean-field spin glasses. Namely, increasing the variance of the random potential $V(\mathbf{x})$ generates an abrupt transition to a phase with exponentially many local minima and saddle-points of $L(\textbf{x})$. 
Those features dominate long-time gradient descent \cite{KurchanLaloux} and their statistics 
has been subject of steady interest in recent years in that and related models \cite{Fyo04,BD07,FyoWi07,FyoNad2012,Auf1,Auf2,SubZei15,Sub2017,Ros2018,Ros2020,NBetal,AufZeng}.  As shown in
\cite{FK2016,BFK2020} adding random non-potentiality in the right-hand side of (\ref{model}) has profound effect on
the phase portrait. Namely, increasing interaction strength makes such systems to undergo an abrupt transition to a regime of 'absolute instability' where points of equilibria are on average exponentially abundant, but typically all of them are unstable, unless the dynamics is purely gradient. 
The authors  were able also to calculate the mean proportion of points of equilibria which have a fixed fraction of unstable directions.  Interestingly, the model actually shared the same rate of exponential growth in the total number of equilibria close to the threshold as one in \cite{WT},  pointing towards a certain universality of the predicted scenario beyond the tipping point instability.

   Despite successfully revealing a rich structure underlying the phase portrait beyond the instability threshold, the model considered in \cite{FK2016,BFK2020} has been shown to have almost surely only a single (stable) fixed point globally before the instability threshold develops. Hence, in that parameter regime such models lack any nontrivial phase space structure and provide no room for revealing resilience mechanisms as defined in the introduction. That property seems to be intimately related to choosing the random vector field ${\bf f}(\bm x)$ to be homogeneous, i.e. statistically invariant with respect to spatial translations. Moreover,  the specified choice of ${\bf f}(\bm x)$ made to ensure analytical tractability simultaneously imposed a certain departure from the spirit of the original May's analysis. Namely, as for such a choice ${\bf f}(\bm 0)\ne 0$  almost surely, the origin ${\bf x}=0$ is almost never a point of dynamical equilibrium of the nonlinear system (\ref{model}). In this sense the framework provided by the model of  \cite{FK2016,BFK2020} appears less suitable for addressing the implications of nonlinearity on the (in)stability of a {\it given} equilibrium, replacing the original May's question with statistical analysis of totality of fixed points.  At the same time focusing analysis on mechanisms of building instability in the vicinity of a chosen equilibrium is clearly highly desirable.

 In the present letter we aim to suggesting an alternative mathematical model which allows to address the above questions for the system (\ref{model}) in considerable generality.  To this end we replace the linear interaction term in~\eqref{may} with a full (Taylor) expansion around the fixed point and thereby include nonlinearity through higher-order interactions. It is convenient to normalize
  the interaction strength with $N$ as ${\bf f}(\bm x):=\frac1{\sqrt N}\bm\phi(\bm x)$ which will ensure natural
 behaviour in the large-$N$ limit, and further define $\bm\phi(\bm x)$
 via an expansion
\begin{equation}\label{phi-def}
\phi_n(\bm x)=\sum_{k=1}^\infty\sigma_k\sum_{i_1,\ldots,i_k=1}^N\xi_{n,i_1,\ldots,i_k}x_{i_1}\cdots x_{i_k}
\end{equation}
where $\sigma_k>0$ are positive constants determining the strength of the $k$-order interactions. Our main assumption is that $\xi_\bullet$ are i.i.d. Gaussian variables with zero mean and unit variance, i.e.
\begin{align}
\E[\xi_{n,i_1,\ldots,i_k}]&=0\\
\E[\xi_{n,i_1,\ldots,i_k}\xi_{m,j_1,\ldots,j_\ell}]&=
\delta_{nm}\delta_{k\ell}\delta_{i_1j_1}\cdots\delta_{i_kj_k}.
\end{align}
In the spirit of May's original model, we have chosen the system to be fully randomly assembled, i.e. all interactions (including higher-order terms) are independently distributed. However in contrast to \cite{FK2016,BFK2020} the random interactions in our dynamical system are now not statistically invariant with respect to spatial translations, with the origin always singled out as a fixed point of the dynamics. The latter feature of our model is therefore shared with that in \cite{WT}. At the same time, in contrast to \cite{WT} the nonlinear interactions in our system still retain a highly symmetric statistical nature, and allow to develop a fully controllable (and essentially rigorous) method of its analysis.
In particular, the above given definitions imply the following spatial covariance structure for the random vector field $\bm \phi$:
\begin{equation}\label{mean+corr}
\E[\phi_n(\bm x)]=0,
\qquad
\E[\phi_n(\bm x)\phi_m(\bm y)]=\delta_{nm}C(\bm x^T\bm y)
\end{equation}
with the (scalar) correlation function given by
\begin{equation}\label{C}
C(\bm x^T\bm y)=\sum_{k=1}^\infty \sigma_k^2(\bm x^T\bm y)^{k}.
\end{equation}
where $\E[\ldots]$ stands for the expected value. 
The assumption about Gaussianity is technically convenient since it implies that the random vector field $\bm \phi$ is fully determined by the mean and covariance structure~\eqref{mean+corr}, which allows us to perform explicit and fully controllable derivations of our main results. Nonetheless, it is natural to expect that our main conclusions should hold beyond the Gaussian case under proper assumptions on the higher moments of the random variables $\xi_\bullet$.

In order to allow a nonlocal analysis, we must require that the constants $\sigma_1,\sigma_2,\ldots$ decay sufficiently fast for large $k$ to ensure a nonzero (including infinite) radius of convergence $R\in (0,\infty]$. Examples include $\sigma_k=1/\sqrt{k!}$ which implies $C(z)=e^{z}-1$ and $R=\infty$, and $\sigma_k=1$ which implies $C(z)=z/(z-1)$ and $R=1$. Convergence (in probability) of the random field $\bm\phi(\bm x)$ follows e.g. by Markov's inequality. Henceforth, it is always assumed that $\abs{\bm x}<R$.

We would like to emphasize the following two distinct properties ensured by our choice of the random vector field $\bm \phi(\bm x)$:
\begin{itemize}

\item [(a)] We have the linear approximation $\bm\phi(\bm x)\approx \sigma_1\bm J\bm x$ for small $\bm x$, so the nonlinear model~\eqref{model} preserves May's linear approximation and thereby his instability criteria. In other words, the fixed point at the origin is locally stable for $\mu>\sigma_1$ as predicted by the linear model~\eqref{may}. The inclusion of higher-point interaction terms will allow us to address nonlocal properties within the region of convergence, i.e. for $\abs{\bm x}<R$.

\item [(b)] The random vector field $\bm\phi(\bm x)$ is statistically bi-rotational invariant, that is the mean and correlation function~\eqref{mean+corr} are invariant under the transformation $\bm\phi(\bm x)\mapsto \bm V\bm\phi(\bm U\bm x)$ for all rotations (including improper rotations) $\bm U,\bm V\in O(N)$. Note that in the full system~\eqref{model} with $\mu>0$ this symmetry is explicitly broken down by the stability feedback mechanism from $O(N)\times O(N)$ to $O(N)$,  with bi-rotational symmetry becoming the ordinary rotational symmetry (isotropy), such that we must have $\bm V=\bm U$. 

\end{itemize}

 In any stability analysis of an autonomous dynamical systems, a natural first step is to determine the number and location of fixed points, and then classify them by stability.  
Due to the randomness of the vector field $\bm \phi$, the number of fixed points and their locations will be random too (with the important exception of the origin which remains a fixed point by our construction). To characterize the spatial distribution of fixed points we introduce the mean fixed point density $\rho_\mu(\bm x)$, such that the mean number of fixed points in a domain $D$ is given by
\begin{equation}
\E[\#\{\textup{fixed point in }D\}]=\int_D d\bm x\,\rho_\mu(\bm x).
\end{equation}
Since the origin is a fixed point by construction, the mean spectral density $\rho_\mu(\bm x)$ contains a Dirac delta function with unit mass at the origin, and this point must in general be treated separately. Away from the origin ($\bm x\neq 0)$, the mean density of fixed points $\rho_\mu(\bm x)$ can be analysed using the so-called Kac--Rice method, see e.g.~\cite{AT2009,AW2009,Fyodorov2015} and references therein. To apply the Kac--Rice formula, one needs the joint probability density of the random vector field $\bm \phi$ and its first derivatives. 
It is convenient to represent all the first derivatives as an $N\times N$ matrix-valued random field $\nabla\bm \phi(\bm x)=[\p_n\phi_m(\bm x)]_{n,m}$.
With this notation, the Kac--Rice formula states that 
\begin{equation}\label{kac-rice}
\rho_\mu(\bm x)=\int_{\R^{N\times N}}d\bm M\,p(\mu\sqrt N\bm x,\bm M)
\abs{\det(\bm M-\mu\sqrt N\bm 1_N)},
\end{equation}
where $p(\bm v,\bm M)$ is the joint probability density function (PDF) for $\bm v=\bm\phi(\bm x)$ and $\bm M=\nabla\bm\phi(\bm x)$.
Due to our assumption of Gaussianity, the fields $\bm\phi$ and $\nabla\bm\phi$ are jointly Gaussian and the PDF $p$ can be found using standard techniques, see the Supplemental Material for more detail, as given by 
\begin{equation}\label{pdf}
p(\bm v,\bm M)=
\frac{
e^{-\tfrac{\bm v^T\bm v}{2C}}
e^{-\tfrac{
\tr (\bm M-\frac{C'}{C}\bm v\bm x^T)
\bm S(\bm x)
(\bm M-\frac{C'}{C}\bm v\bm x^T)^T
}{2\Delta C'}}
}
{(2\pi)^{N(N+1)/2}\Delta^{N/2}(C')^{N(N-1)/2}}
\end{equation}
with
\begin{align}
\Delta&=CC'+\big(CC''-(C')^2\big)\bm x^T\bm x,\\
\bm S(\bm x)&=
\Delta\bm 1_N-(CC''-(C')^2)\bm x\bm x^T,
\end{align}
and $C$ denoting the scalar correlation function~\eqref{C}, while $C'$ and $C''$ stands for its first and second derivative with respect to its scalar argument. Here, we have suppressed the explicit dependence on $\bm x^T\bm x$ for $C,C',C''$ and $\Delta$.
We observe that as a consequence of the bi-rotational invariance of the random vector field $\bm\phi$, the PDF~\eqref{pdf} is invariant under the transformation
\begin{equation}
\big(\bm x,\bm v,\bm M\big)
\mapsto
\big(\bm U\bm x,\bm V\bm v,\bm V\bm M\bm U^T\big)
\end{equation}
for all rotations $\bm V,\bm U\in O(N)$.
In fact, up to the scalar functions of $\bm x^T\bm x$, the form of the PDF~\eqref{pdf} is fully determined by this symmetry together with assumption of centred Gaussianity.

Upon inserting the  expression~\eqref{pdf} into the Kac--Rice formula~\eqref{kac-rice} and making a change of variables
\begin{equation}\label{cov}
\bm M\mapsto \bm M\Big(\frac{\bm S(\bm x)}{\Delta C'}\Big)^{-1/2}+\mu\sqrt N\frac{C'}{C}\bm x\bm x^T
\end{equation}
one is able to considerably simplify \eqref{kac-rice} and bring it to the form
\begin{multline}\label{kac-rice-full}
\rho_\mu(\bm x\neq 0)=
\frac{1}{(2\pi)^{N/2}}\Big(\frac\Delta{CC'}\Big)^{\frac12}\Big(\frac{C'}{C}\Big)^{\frac{N}2}
e^{-{N\mu^2\bm x^T\bm x}/{2C}}\\
\times\E_\text{Gin}\big[\abs{\det(\bm \varXi-\mu\sqrt N\bm D)}\big],
\end{multline}
where
\begin{equation}\label{deform}
\bm D=\diag\bigg\{\sqrt{\frac {C}\Delta}\Big(1-\frac{C'}{C}\bm x^T\bm x\Big),\sqrt{\frac1{C'}},\ldots,\sqrt{\frac1{C'}}\bigg\}
\end{equation}
is a diagonal matrix
and $\E_\text{Gin}$ denotes the expectation taken with respect to $N\times N$ random matrices from the so-called real Ginibre ensemble \cite{KhorSomRev}, with all entries being i.i.d. standard real mean-zero Gaussian variables. 
Note that the matrix $\bm S(\bm x)/\Delta C'$ is positive definite symmetric for $\bm x\neq 0$, so that the transformation~\eqref{cov} is well-defined.


So far our analysis has been exact for any finite number of interacting degrees of freedom $N$, but our main concern is for large systems when $N\to\infty$. Before being able to extract the large-$N$ asymptotic behaviour of the  density ~\eqref{kac-rice-full}, we have to recall some basic properties of the (scalar) correlation function $C(r^2)$ and introduce notations convenient for achieving this goal.

Using the definition of the correlation function $C$, it is straightforward to verify that
$C'(r^2)$ and ${C(r^2)}/{r^2}$ are continuous and strictly monotonically increasing for $0<r<R$ and that they tend to $\sigma_1^2>0$ for $r\to0^+$ and to infinity as $r\to R$. The ratio and difference combinations defined as ${C'(r^2)r^2}/{C(r^2)}$ and $C'(r^2)-{C(r^2)}/{r^2}$ are also continuous and strictly monotonically increasing for $0<r<R$. We will use these monotonicity properties frequently in the following analysis.

Now, we can introduce two new radii $r_\pm(\mu)\geq 0$ in the following way: we set
$r_\pm(\mu)=0$ as long as $\mu\leq\sigma_1$, whereas
 for $\mu>\sigma_1$ the two radii are defined as the solutions to the equations
$C'(r_-^2)=\mu^2$ and ${C(r_+^2)}=\mu^2{r_+^2}$, respectively. From the monotonicity properties described above, we know that $r_\pm(\mu)$ are uniquely defined, continuous and monotonically increasing as functions of $\mu$. Furthermore, we have
$0<r_-(\mu)<r_+(\mu)$ for all $\mu>\sigma_1$, with $r_\pm(\mu)\to0$ for $\mu\to\sigma_1^+$ and $r_\pm(\mu)\to R$ for $\mu\to\infty$.

With this new notations at hand, we can embark on our large-$N$ analysis. For this it is convenient to define the mean `spherical' density of the fixed points as
\begin{equation}\label{rho-hat}
\widehat\rho_\mu(r>0):=\frac{2\pi^{N/2}r^{N-1}}{\Gamma(N/2)}\rho_\mu(r).
\end{equation}
Here, the prefactor in the right-hand side is the surface area of an $(N-1)$-dimensional sphere of radius $r$.
We will see that there exists a characteristic radius $r_*(\mu)\in [r_-(\mu),r_+(\mu)]$ such that the spherical mean density of fixed points is low for $r<r_*$ and high for $r>r_*$.

The main challenge in extracting the large-$N$ asymptotics of the spherical density~\eqref{rho-hat} is in performing an asymptotic evaluation of the expectation value in the second line in~\eqref{kac-rice-full}. Evaluating averages involving absolute values of determinants of random matrices and operators is a common step in counting problems  based on the Kac-Rice method and as such attracted considerable attention in recent years, see e.g.  \cite{IF2018,FK2016,Fyo04,FyoWi07,Auf1,Auf2,FLRTa,Ros2018,NBetal,AufZeng}. Here the challenge is related to performing such a calculation
for the real Ginibre ensemble deformed by a finite-rank perturbation, see (\ref{deform}).
Fortunately, it turns out that the object in question can be evaluated rigorously by adapting the approach suggested in~\cite{Fyodorov2018a} for a rather different problem. Relegating the detail of the calculation to the Supplemental Material, we give below the ensuing expression for the mean (spherical) density of fixed points:
\begin{equation}\label{rho-asymp-}
\widehat\rho_\mu(r)=\sqrt{\frac N\pi}\frac{h_\textup{I}(r^2)}r e^{+\frac N2 L_\textup{I}(r^2)}(1+o(1))
\end{equation}
for $0<r<r_-$ and
\begin{equation}\label{rho-asymp+}
\widehat\rho_\mu(r)=\sqrt{\frac N\pi}\frac{h_\textup{II}(r^2)}r e^{+\frac N2 L_\textup{II}(r^2)}(1+o(1))
\end{equation}
for $r_-<r<R$, where
\begin{align}
h_\textup{I}(r^2)&=\tfrac{C'}{C}r^2-1, \\
h_\textup{II}(r^2)&=\big(\tfrac{2\Delta}{CC'}\big)^{\frac12}\big(1+\mu^2\big(\tfrac{C}{\Delta}h_\textup{I}(r^2)^2-\tfrac{1}{C'}\big)\big)^{\frac12}.
\end{align}
and
\begin{align}
L_\textup{I}(r^2)=-f\big(\tfrac{\mu^2r^2}{C}\big),\quad
L_\textup{II}(r^2)=f\big(\tfrac{\mu^2}{C'}\big)-f\big(\tfrac{\mu^2r^2}{C}\big)
\end{align}
with $f(x)=x-\log x-1$ ($x>0$).

Note that~\eqref{rho-asymp-} and \eqref{rho-asymp+} represent a density, thus these expressions must be nonnegative.
It follows that the functions $h_\textup{I}(r^2)$ and $h_\textup{II}(r^2)$ must be nonnegative in their relative domains (i.e. for $0<r<r_-$ and $r_-<r<R$, respectively), which can be indeed verified using the monotonicity properties of scalar correlation function.

Our next task is to investigate for which values of the radius $r$ the exponent $L_\bullet$ is positive or negative, implying the spherical density of fixed points being, respectively, exponentially suppressed or enhanced. We will consider three different intervals separately:

\begin{itemize}
\item[(\textit{i})] For $0<r<r_-$, it is easily verified that $L_\textup{I}(r^2)<0$. Moreover, $L_\textup{I}(r^2)$ is monotonically increasing on this interval.
\item[(\textit{ii})]
For $r_-<r<r_+$, we note that $L_\textup{II}(r^2_-)<0<L_\textup{II}(r^2_+)$, so $L_\textup{II}(r^2)$ vanishes at least once in this interval. By differentiation, we have
\begin{align}\label{LII'}
\qquad L_\textup{II}'(r^2)=
\frac{(C'-\mu^2)C''}{(C')^2}+h_\textup{I}(r^2)\Big(\mu^2-\frac{C}{r^2}\Big).
\end{align}
Using the monotonicity properties of the correlation function, we see that the first term on the right-hand side in~\eqref{LII'} is positive for $r_-<r<R$ and that second term on the right-hand side in~\eqref{LII'} is positive for $0<r<r_+$. Thus, $L_\textup{II}(r^2)$ is strictly monotonically increasing for $r_-<r<r_+$ and there exists a unique radius $r_*\in[r_-,r_+]$ such that $L_\textup{II}(r^2)<0$ for $r<r_*$ and $L_\textup{II}(r^2)>0$ for $r>r_*$.
\item[(\textit{iii})]
For $r_+<r<R$, we can use the following properties of the function $f$: we have $f(x)-f(y)>0$ for $x<y<1$ and $f(x_2)-f(y_2)>f(x_1)-f(y_1)$ for $0<y_2-x_2<y_1-x_1$ and $x_2/y_2<x_1/y_1<1$. It follows that
$L_\textup{II}(r^2)$ is strictly positive and monotonically increasing.
\end{itemize}

By definition, the mean number of fixed points within a ball of radius $0<r<R$ centred at the origin is given by
\begin{equation}\label{ball}
\mathcal N_\mu(r)
=\int_{\abs{\bm x}<r}d\bm x\,\rho_\mu(\bm x)
=1+\int_{0<\tilde r<r}d\tilde r\,\widehat\rho_\mu(\tilde r),
\end{equation}
where we separated in the final expression the result of integration over the Dirac mass at the origin.
It is straightforward to use the properties (\textit{i}-\textit{iii}) above to give exponential bounds for this quantity. We see that there exists strictly positive functions $c_1(r),c_2(r),\kappa_1(r),\kappa_2(r)$ (independent of $N$) such that
\begin{align}
\mathcal N_\mu(r)-1&\leq ({\tfrac N\pi})^{\frac12} c_1(r)e^{-N \kappa_1(r)}&& \text{for}\! & r&\in(0,r_*), \label{ineq-small} \\
\mathcal N_\mu(r)-1&\geq ({\tfrac N\pi})^{\frac12} c_2(r)e^{+N\kappa_2(r)}&& \text{for}\! & r&\in(r_*,R). \label{ineq-large}
\end{align}
As an example, one possible choice for the functions $c_1(r),c_2(r),\kappa_1(r),\kappa_2(r)$ is
\begin{align}
c_1(r)&=
\begin{cases}
\int_0^r\frac{d\tilde r}{\tilde r}{h_\textup{I}(\tilde r^2)}, & r<r_-\\
\int_0^{r_-}\frac{d\tilde r}{\tilde r}{h_\textup{I}(\tilde r^2)}
+\int_{r_-}^r\frac{d\tilde r}{\tilde r}{h_\textup{II}(\tilde r^2)}, & r>r_-
\end{cases},\\
c_2(r)&=\int_{r_*}^r\frac{d\tilde r}{\tilde r}{h_\textup{II}(\tilde r^2)},\\
\kappa_1(r)&=
\begin{cases}
-L_\textup{I}(r^2)/2, & r<r_-\\
-L_\textup{II}(r^2)/2, & r>r_-
\end{cases},\\
\kappa_2(r)&=+L_\textup{II}(r^2)/2.
\end{align}
Here, the positivity of the functions $c_1$ and $c_2$ is evident from positivity of the functions $h_\textup{I}$ and $h_\textup{II}$, while positivity of $\kappa_1$ and $\kappa_2$ follows from properties (\textit{i}-\textit{iii}).

The inequalities~\eqref{ineq-small} and~\eqref{ineq-large} tell us that the mean number $\mathcal N_\mu(r)-1$  of fixed points in a ball centred around the fixed point at the origin, but different from it, is exponentially suppressed for large $N$ when the radius of the ball satisfies $r<r_*(\mu)$.
In fact, it implies that the origin is almost surely (with probability 1) the only fixed point within a ball with radius $r<r_*(\mu)$ for $N\to\infty$, since the mean value equals the lowest possible value.
In contrast, inside any ball whose radius exceeds the threshold value
$r=r_*(\mu)$, the mean number of fixed points effectively generated by
nonlinear couplings grows exponentially fast with increasing $N$.
We note that even though there are exponentially many fixed points beyond the critical radius $r_*$, the mean fixed point density~\eqref{kac-rice-full} itself can be small at a location $\bm x$ with $\abs{\bm x}>r_*$.

In order to interpret this result, let us first consider a dynamical system with $\mu>\sigma_1$, in which case May's linear approximation tells us that the fixed point at the origin is locally stable with probability one. Including higher order random terms in the expansion around the fixed point thus extends this local stability result to a global one by telling us that there is a certain {\it resilience gap} surrounding the locally stable origin. Namely, no other fixed points can be almost surely found within a radius $r_*(\mu)>0$, but exponentially many fixed point exist beyond this radius. Based on that it is natural to suggest that $r_*(\mu)$ should play the role of a {\it resilience radius}, a characteristic scale of  sensitivity of system's behaviour to initial displacement. Namely, trajectories corresponding to initial conditions such that $|{\bf x}(0)|\lesssim r_*(\mu)$ are naturally expected to be typically attracted to the stable origin, but those which start with $|{\bf x}(0)|\gtrsim r_*(\mu)$ might leave the origin's basin of attraction to wander away in the maze of exponentially many fixed points, and possibly become eventually chaotic. Although this picture at the moment remains largely speculative, and certainly requires further investigation, our rigorous results provide a basis for suggesting it as a possible mechanism of nonlinearity-generated resilience for large complex systems. It is worth emphasizing that  the existence of the resilience gap in the present model is quite universal: it does not depend on the choice of $\sigma_2,\sigma_3,...$ as long as they are not all zero.

 If one takes $\mu\to \infty$, then by using $r_-(\mu)\leq r_*(\mu)\leq r_+(\mu)$ and $r_-(\mu),r_+(\mu)\to R$,
we conclude that the resilience radius takes the largest allowed value: $r_*(\mu)\to R$. This implies that with growing $\mu$ all other fixed points are  pushed away from the  stable origin to the radius of convergence of the Taylor expansion (which might be infinity), so that the system becomes more and more resilient to initial displacements. On the other hand, if we take $\mu\to\sigma_1$ (from above) then $r_*(\mu)\to 0$, hence the system become less and less resilient to perturbations in initial conditions, and trajectories starting relatively close to the origin are expected to wander away. Finally, when the linear stability threshold/tipping point is crossed (i.e. $\mu<\sigma_1$) the fixed point at the origin becomes locally unstable and any ball (with positive radius) centred at the origin will contain on average exponentially many other fixed points. Thus, we expect dynamics in this `unstable regime' to be extremely sensitive to initial displacements. There are plenty of questions which need to be clarified for the present model, most immediate is
to attempt a classification of the fixed points away from the origin by their stability properties, not unlike the analysis recently performed in  \cite{BFK2020}. More generally, an accurate analysis of long-time autonomous dynamics based on the system of randomly coupled ODE's (\ref{model}) remains largely open and poses an outstanding challenge.
Note that not unrelated recent studies of species dynamics in the framework of generalized Lotka-Volterra models revealed many intriguing features, such as marginal stability of the ensuing equilibria, see \cite{BBC18} and references therein.
Another interesting avenue to explore is the effect of time-dependent landscapes~\cite{IS2016,Ip2017}.



\pagebreak

~

~

~

\widetext
\begin{center}
\textbf{\large Supplemental Materials for\\ ``Nonlinearity-generated Resilience in Large Complex Systems''}
\end{center}

\setcounter{equation}{0}
\setcounter{figure}{0}
\setcounter{table}{0}
\setcounter{page}{1}
\makeatletter
\renewcommand{\theequation}{S\arabic{equation}}
\renewcommand{\thefigure}{S\arabic{figure}}
\renewcommand{\bibnumfmt}[1]{[S#1]}
\renewcommand{\citenumfont}[1]{S#1}
\makeatother

\section{Derivation of the joint probability density function}

The goal of this section is to establish the joint PDF for the random vector $\bm \phi(\bm x)$ and the random matrix $\nabla\bm \phi(\bm x)$. We know from the definition of the random vector field that $\bm \phi(\bm x)$ and  $\nabla\bm \phi(\bm x)$ are jointly Gaussian. Furthermore, it follows by differentiating of the correlation function that both fields are mean zero and their covariances at $\bm x$ reads
\begin{align}
\E[\phi_n(\bm x)\phi_m(\bm x)]&=\delta_{nm}C(\bm x^T \bm x), \label{var1}\\
\E[\p_k\phi_n(\bm x)\phi_m(\bm x)]&=\delta_{nm}x_kC'(\bm x^T \bm x), \label{var2}\\
\E[\p_k\phi_n(\bm x)\p_\ell\phi_m(\bm x)]&=\delta_{nm}\delta_{k\ell}C'(\bm x^T \bm x)+\delta_{nm}x_kx_\ell C''(\bm x^T \bm x). \label{var3}
\end{align}
By the standard theory of multivariate Gaussians, we know that the joint distribution of $\bm \phi(\bm x)$ and the random matrix $\nabla\bm \phi(\bm x)$
\begin{equation}
p(\bm v,\bm M)= \frac{\prob[\bm\phi(\bm x)\in (\bm v,\bm v+d\bm v),\nabla\bm\phi(\bm x)\in (\bm M,\bm M+d\bm M)]}{d\bm v\, d\bm M}.
\end{equation}
is given by
\begin{equation}\label{pdf-vector}
p(\bm v,\bm M)=
\frac1{(2\pi)^{N(N+1)/2}(\det\bm\Sigma(\bm x))^{1/2}}\exp\bigg(-\frac12
\vec[\bm v,\bm M]^T
\bm\Sigma^{-1}(\bm x)
\vec[\bm v,\bm M]
\bigg),
\end{equation}
where `$\vec$' is the vectorisation operator which turns a matrix into a column vector by stacking its columns on top of each other, i.e. $\vec[\bm v,\bm M]$ is an $N(N+1)$ column vector. The covariance matrix $\bm \Sigma(\bm x)$ is an $N(N+1)\times N(N+1)$ matrix defined by
\begin{equation}
\bm\Sigma(\bm x)=\E[\vec[\bm v,\bm M]\vec[\bm v,\bm M]^T]
\end{equation}
It follows from~\eqref{var1}, \eqref{var2}, and~\eqref{var3} that
\begin{equation}\label{Sigma=sigma}
\bm\Sigma(\bm x)=
\bm\sigma(\bm x)\otimes \bm 1_N,
\end{equation}
where $\otimes$ denotes the Kronecker (tensor) product, $\bm 1_N$ is the $N\times N$ identity matrix, and
\begin{equation}
\bm\sigma(\bm x)=
\begin{bmatrix}
C(\bm x^T\bm x) & C'(\bm x^T\bm x)\bm x^T \\
C'(\bm x^T\bm x)\bm x & C'(\bm x^T\bm x)\bm 1_N+C''(\bm x^T\bm x)\bm x\bm x^T
\end{bmatrix}
\end{equation}
is an $(N+1)\times (N+1)$ matrix.

In writing down the PDF~\eqref{pdf-vector}, we have assumed that the covariance $\bm\Sigma(\bm x)$ is invertible. Before we continue, we must verify this fact. We have
\begin{equation}
\det[\bm\Sigma(\bm x)]=\det[\bm\sigma(\bm x)\otimes \bm 1_N]=\det[\bm\sigma(\bm x)]^N,
\end{equation}
thus we need to verify that the determinant of $\bm\sigma(\bm x)$ is nonzero. In order to evaluate this determinant, we first recall the standard block matrix identity for determinants
\begin{equation}
\det
\begin{bmatrix}
\bm A & \bm B \\
\bm C & \bm D
\end{bmatrix}
=\det[\bm A]\det[\bm D-\bm C\bm A^{-1}\bm B]
\end{equation}
with $\bm A$ an invertible $n\times n$ matrix, and $\bm B,\bm C,\bm D$ any $n\times m$, $m\times n$, $m\times m$ matrices. Using this identity, we get
\begin{equation}\label{det_Sigma}
\det[\bm\sigma(\bm x)]=
C(\bm x^T\bm x)C'(\bm x^T\bm x)^N
\det\Big[\bm 1_N+\frac{C(\bm x^T\bm x)C''(\bm x^T\bm x)-C'(\bm x^T\bm x)^2}{C(\bm x^T\bm x)C'(\bm x^T\bm x)}\bm x\bm x^T\Big].
\end{equation}
To evaluate the remaining determinant in~\eqref{det_Sigma}, we use another determinant identity. We have
\begin{equation}
\det[\bm 1_n+\bm A\bm B]=\det[\bm 1_m+\bm B\bm A]
\end{equation}
for matrices $\bm A$ and $\bm B$ of size $n\times m$ and $m\times n$, respectively. Thus, determinant reads
\begin{equation}\label{det_Sigma2}
\det[\bm\sigma(\bm x)]=\Delta(\bm x^T\bm x)C'(\bm x^T\bm x)^{N-1}.
\end{equation}
where we have introduced the scalar function
\begin{equation}\label{Delta}
\Delta(\bm x^T\bm x)=C(\bm x^T\bm x)C'(\bm x^T\bm x)+\big(C(\bm x^T\bm x)C''(\bm x^T\bm x)-C'(\bm x^T\bm x)^2\big)\bm x^T\bm x
\end{equation}
for notational simplicity. We note that the determinant~\eqref{det_Sigma2} depends only on the squared (Euclidean) distance to the origin $r^2=\bm x^T\bm x\geq 0$.

Our first observation about the determinant~\eqref{det_Sigma2} is that it equals zero for $\bm x=0$, hence the covariance matrix $\bm \Sigma(\bm x)$ is not invertible for $\bm x=0$. This arises from the fact that $\bm\phi(0)=0$ and, thus, non-random.
For $\bm x\neq 0$, the vector field $\bm \phi$ is truly random. In this case, we can verify from the definition of the correlation function $C$ that $C'(r^2)$ and $\Delta(r^2)$ are positive (and finite) for $0<r<R$ (both $C'(r^2)$ and $\Delta(r^2)$ blow up when $r\to\R$). Consequently, the determinant~\eqref{det_Sigma2} is non-zero from which it follows that the covariance matrix $\bm\Sigma(\bm x)$ is invertible and thereby that PDF~\eqref{pdf-vector} is valid for all $0\leq\abs{\bm x}\leq R$.

Now that we have verified that the convariance matrix $\bm \Sigma (\bm x)$ is indeed invertible for $\bm x\neq0$, let us find its inverse. We have
\begin{equation}\label{Sigma-1=sigma-1}
\bm \Sigma (\bm x)^{-1}=\bm\sigma (\bm x)^{-1}\otimes \bm 1_N
\end{equation}
for $\bm x\neq 0$. We recall the block inverse identity
\begin{equation}
\begin{bmatrix}
\bm A & \bm B \\
\bm C & \bm D
\end{bmatrix}^{-1}
=
\begin{bmatrix}
\bm A^{-1}+\bm A^{-1}\bm B(\bm D-\bm C\bm A^{-1}\bm B)^{-1}\bm C\bm A^{-1} & -\bm A^{-1}\bm B(\bm D-\bm C\bm A^{-1}\bm B)^{-1} \\
-(\bm D-\bm C\bm A^{-1}\bm B)^{-1}\bm C\bm A^{-1} & (\bm D-\bm C\bm A^{-1}\bm B)^{-1}
\end{bmatrix}
\end{equation}
where $\bm A ,\bm B ,\bm C, \bm D$ are $n\times n$, $n\times m$, $m\times n$, $m\times m$ matrices with $\bm A$ and $(\bm D-\bm C\bm A^{-1}\bm B)$ invertible.
Using this identity, a straightforward computation yields
\begin{equation}
\bm\sigma(\bm x)^{-1}=
\frac{1}{\Delta(\bm x^T\bm x)C'(\bm x^T\bm x)}
\begin{bmatrix}
C'(\bm x^T\bm x)^2+C'(\bm x^T\bm x)C''(\bm x^T\bm x)\bm x^T\bm x & -C'(\bm x^T\bm x)^2\bm x^T \\
-C'(\bm x^T\bm x)^2\bm x & \bm S(\bm x)
\end{bmatrix}
\end{equation}
with $\Delta(\bm x^T\bm x)$ given by~\eqref{Delta} and
\begin{equation}\label{S}
\bm S(\bm x)=\Delta(\bm x^T\bm x)\bm 1_N-(C(\bm x^T\bm x)C''(\bm x^T\bm x)-C'(\bm x^T\bm x)^2)\bm x\bm x^T
\end{equation}
a symmetric matrix-valued function.

We now have explicit expressions for all quantities which appears in~\eqref{pdf-vector}, thus we know the full distribution.
However, we still want to reexpress the PDF as a matrix Gaussian distribution rather than the standard multivariate form.
We recall the following identity involving the Kronecker product and the vectorisation operator
\begin{equation}
\bm A\otimes \bm B\vec[\bm X]=\vec[\bm B\bm X\bm A^T]
\end{equation}
for matrices $\bm A,\bm B,\bm X$ of size $k\times m$, $n\times m$, $\ell\times n$. Using this identity and the fact that $\bm\sigma(\bm x)$ is a symmetric matrix, we see that
\begin{equation}
\vec[\bm v,\bm M]^T\bm \Sigma(\bm x)^{-1}\vec[\bm v,\bm M]=
\tr[\bm v,\bm M]\bm\sigma(\bm x)^{-1}[\bm v,\bm M]^T,
\end{equation}
where $[\bm v,\bm M]$ is an $N\times(N+1)$ matrix. Thus, the PDF~\eqref{pdf-vector} reads
\begin{equation}
p(\bm v,\bm M)=
\frac1{(2\pi)^{N(N+1)/2}(\det\bm\sigma(\bm x))^{N/2}}
\exp\Big(-\frac12\tr[\bm v,\bm M]\bm\sigma(\bm x)^{-1}[\bm v,\bm M]^T\Big).
\end{equation}
Expanding in terms of $\bm v$ and $\bm M$ and completing the square yields
\begin{multline}
p(\bm v,\bm M)=
\frac1{(2\pi)^{N(N+1)/2}\Delta(\bm x^T\bm x)^{N/2}C'(\bm x^T\bm x)^{N(N-1)/2}}\\
\times\exp\bigg[
-\frac{\bm v^T\bm v}{2C(\bm x^T\bm x)}-
\frac1{2\Delta(\bm x^T\bm x)C'(\bm x^T\bm x)}\tr
\Big(\bm M-\frac{C'(\bm x^T\bm x)}{C(\bm x^T\bm x)}\bm v\bm x^T\Big)
\bm S(\bm x)
\Big(\bm M-\frac{C'(\bm x^T\bm x)}{C(\bm x^T\bm x)}\bm v\bm x^T\Big)^T\bigg],
\end{multline}
which is the expression that we wanted to establish.

\section{Evaluation of the matrix average}

The purpose of this section is to reexpress matrix average that appear on the second line in~\eqref{kac-rice-full} as an expression which is more suitable for a large-$N$ asymptotic analysis (our approach similar to the approached used in~\cite{Fyodorov2018,Fyodorov2019}). Let us consider the generalised problem
\begin{equation}\label{gen-average}
\E_\text{Gin}[|\det(\lambda\bm 1_N+\epsilon\bm h \bm h^T-\bm \varXi)|]
\end{equation}
for $\lambda, \varepsilon\in \R$ and $\bm h\in \R^N$. The required expectation in (\ref{kac-rice-full}) is  retrieved by imposing
\begin{equation}\label{parameter-choice}
\lambda=\mu\sqrt{\frac{N}{C'(r^2)}},\qquad
\varepsilon=\mu\sqrt{N\frac{C(r^2)}{\Delta(r^2)}}\Big(1-\frac{C'(r^2)}{C(r^2)}r^2\Big)-\mu\sqrt{\frac{N}{C'(r^2)}},\qquad\text{and}\qquad
\bm h=(1,0,...,0)^T.
\end{equation}
In order to derive an explicit expression for the average~\eqref{gen-average}, we make two important observations. First, we observe that for any real $x$, the absolute value $|x|$ can be written as $x^2/\sqrt{x^2}$, which allows us to write the average~\eqref{gen-average} as
\begin{equation}
\E_\text{Gin}[|\det(\bm\Lambda-\bm \varXi)|]\propto
\E_\text{Gin}
\Bigg[\frac{\det^2(\bm\Lambda-\bm \varXi)}{\sqrt{\det^2(\bm\Lambda-\bm\varXi)}}\Bigg]=
\E_\text{Gin}
\Bigg[\frac{\det\Big(\begin{smallmatrix}0 & i(\bm\Lambda-\bm \varXi) \\ i(\bm\Lambda-\bm \varXi)^T & 0\end{smallmatrix}\Big)\ \ \ }
{\det\Big(\begin{smallmatrix}0 & i(\bm\Lambda-\bm \varXi) \\ i(\bm\Lambda-\bm \varXi)^T & 0\end{smallmatrix}\Big)^{1/2}}\Bigg].
\end{equation}
Here, we have used the shorthand notation $\bm\Lambda=\lambda\bm 1_N+\epsilon\bm h \bm h^T$. Second, we observe that the determinant of any $N\times N$ matrix, $\bm A$, can be written as Berezin integral
\begin{equation}
\det\bm A=\int d\bm \psi d\bm{\tilde{\psi}}\exp{[\bm{\tilde \psi}^T\bm A\bm \psi]}
\end{equation}
where integration is over anti-commuting $N$-dimensional Grassmann variables $\bm \psi$ and $\tilde{\bm\psi}$. Likewise, the reciprocal of the square root of the determinant of a matrix $\bm A$ can be written as an ordinary Gaussian integral
\begin{equation}\label{GausId}
(\det\bm A)^{-1/2}=\frac{1}{(2\pi)^{\frac{N}{2}}}\int_{\R^N} d\bm x \exp{\Big[-\frac{1}{2}\bm x^T\bm A\bm x\Big]}.
\end{equation}
In order to ensure convergence of the integral in~\eqref{GausId}, it is assumed that the real part of the eigenvalues of $\bm A$ are positive . Introducing a regularisation parameter $p\in\R^{+}$, we  can write the expectation as a product of a Berezin and a Gaussian integral
\begin{multline}\label{initial}
\mathcal{D}(\lambda,\varepsilon,\bm h,p)
=\E_\text{Gin}\Bigg[
\frac{\det^2(\bm\Lambda-\bm \varXi)}{\sqrt{\det(2p\bm 1_N+(\bm\Lambda-\bm \varXi)^T(\bm\Lambda-\bm\varXi))}}
\Bigg]
\propto\\
\E_\text{Gin}\Big[\int d\bm\psi d\bm{\tilde{\psi}}e^{-i
\begin{bmatrix}
\tilde{\bm \psi}_1\\
\tilde{\bm \psi}_2
\end{bmatrix}^T
\begin{bmatrix}
\bm 0_N &(\bm\Lambda-\bm \varXi)\\
(\bm\Lambda-\bm \varXi)^T &\bm 0_N
\end{bmatrix}
\begin{bmatrix}
\bm \psi_1\\
\bm \psi_2
\end{bmatrix}}
\int_{\R^{2N}} d\bm{x}_1 d\bm{x}_2 e^{-\frac{1}{2}
\begin{bmatrix}
\bm x_1\\
\bm x_2
\end{bmatrix}^T
\begin{bmatrix}
\sqrt{2p}\bm 1_N &i(\bm\Lambda-\bm \varXi)\\
i(\bm\Lambda-\bm \varXi)^T &\sqrt{2p}\bm 1_N
\end{bmatrix}
\begin{bmatrix}
\bm x_1\\
\bm x_2
\end{bmatrix}}\Big].
\end{multline}
Following the approach contained in~\cite{Fyodorov2018}, the proportionality constant in (\ref{initial}) will be determined later by taking $p\rightarrow+\infty$ while the requested expectation is recovered by taking $p\rightarrow0^{+}$.
We recall that expectation is with respect to real Ginibre matrices, i.e. the entries of $\bm \varXi$ are i.i.d. standard Gaussian random variables.

The argument at the exponent in (\ref{initial}) is linear in $\bm \varXi$ and it can be re-written in term of traces, namely as
\begin{equation}
-\frac{\sqrt{2p}}{2}\tr(\bm x_1\bm x_1^T+\bm x_2\bm x_2^T)-i \bm x_2^T\bm\Lambda\bm x_1+i\tr\bm\Lambda(\bm\psi_2\bm{\tilde\psi_1}^T+\bm\psi_1\bm{\tilde\psi_2}^T)+i\tr\bm \varXi^T(\frac{1}{2}\bm x_1\bm x_2^T-\bm\psi_1\bm{\tilde\psi_2}^T)+i\tr \bm \varXi(\frac{1}{2}\bm x_2\bm x_1^T-\bm\psi_2\bm{\tilde\psi_1}^T).
\end{equation}
The last two terms are integrated out in $\bm \varXi$ by using the following identity for real Ginibre matrices,
\begin{equation}
\E_\text{Gin}\big[e^{-\tr(\bm \varXi\bm A+\bm \varXi^T\bm B)}\big]=e^{\frac{1}{2}\tr(\bm A^T\bm A+\bm B^T\bm B+2\bm A\bm B)}
\end{equation}
Therefore, after introducing a new complex integration variables, $q$ and its complex conjugate $\bar{q}$, in order to recast the nonlinear term $(\tilde{\bm\psi_1}^T\bm\psi_1)(\tilde{\bm\psi_2}^T\bm\psi_2)$ and integrate over the anti-commuting variables, we are left with
\begin{multline}\label{Dd}
\mathcal{D}(\lambda,\varepsilon,\bm h,p)\propto\int_{\C} dq d\bar{q} e^{-|q|^2}\int_{\R^{2N}} d\bm{x}_1 d\bm{x}_2\\
\times \exp\Big[{-\frac{\sqrt{2p}}{2}(\bm x_1^T\bm x_1+\bm x_2^T\bm x_2)-i\bm x_2^T\Lambda\bm x_1-\frac{1}{2}(\bm x_1^T\bm x_1)(\bm x_2^T\bm x_2)}\Big]
\det \begin{bmatrix}
q\bm 1_N &i\bm\Lambda+\bm x_1\bm x_2^T\\
i\bm\Lambda+\bm x_2\bm x_1^T &\bar{q}\bm 1_N
\end{bmatrix}
\end{multline}
A slightly tedious but straightforward computation gives following evaluation of the determinant
\begin{equation}\label{det}
\det \begin{bmatrix}
q\bm 1_N &i\bm\Lambda+\bm x_1\bm x_2^T\\
i\bm\Lambda+\bm x_2\bm x_1^T &\bar{q}\bm 1_N
\end{bmatrix}=(|q|^2+\lambda^2)^{N-3}((|q|^2+\lambda^2)^3-a_2(|q|^2+\lambda^2)^2+a_1(|q|^2+\lambda^2)+a_0)
\end{equation}
with
\begin{align}
a_2=&{}((\bm{x}_1^T\bm x_1)(\bm{x}_2^T\bm x_2)+2 i\varepsilon(\bm x_1^T\bm h)(\bm x_2^T\bm h)-\varepsilon^2(\bm h^T\bm h)^2+2 i\lambda (\bm x_1^T\bm x_2)-2\varepsilon\lambda(\bm h^T\bm h))\\
a_1=&{}-\varepsilon^2((\bm x_1^T\bm h)^2-(\bm h^T\bm h)(\bm x_1^T\bm x_1))((\bm x_2^T\bm h)^2-(\bm h^T\bm h)(\bm x_2^T\bm x_2))+2\varepsilon\lambda (-(\bm x_1^T \bm h)(\bm x_2^T \bm h)(\bm x_1^T \bm x_2) \nonumber \\
&+(\bm x_1^T \bm x_1)(\bm x_2^T \bm h)^2+(\bm x_2^T \bm x_2)(\bm x_1^T\bm h)^2-(\bm h^T\bm h)(\bm x_1^T \bm x_1)(\bm x_2^T \bm x_2)+i \varepsilon (\bm h^T\bm h)((\bm x_1^T \bm h)(\bm x_2^T \bm h) \nonumber \\
&-(\bm h^T\bm h)(\bm x_1^T \bm x_2)))+\lambda^2((\bm x_1^T\bm x_1)(\bm x_2^T\bm x_2)-(\bm x_1^T \bm x_2)^2-4 i\varepsilon(\bm h^T \bm h)(\bm x_1^T \bm x_2)+4i\varepsilon(\bm x_1^T\bm h)(\bm x_2^T\bm h)\\
a_0=&{}\varepsilon(-2(\bm x_1^T\bm h)(\bm x_1^T\bm x_2)(\bm x_2^T\bm h)+(\bm x_1^T\bm x_1)(\bm x_2^T\bm h)^2+(\bm x_2^T\bm x_2)(\bm x_1^T \bm h)^2+(\bm h^T\bm h)((\bm x_1^T\bm x_2)^2 \nonumber\\
&-(\bm x_1^T\bm x_1)(\bm x_2^T\bm x_2)))(-\varepsilon(\bm h^T\bm h)\lambda^2-2\lambda^3)
\end{align}
The integral over the $q$ can be readily solved by introducing the incomplete gamma function $\frac{1}{4\pi}\int d^2q e^{-|q|^2}(|q|^2+\lambda^2)^n=e^{\lambda^2}\Gamma(n+1,\lambda^2)$ for $n\ge0$. After the integration over $q$, the integrand in (\ref{Dd}), can be written as function of a $2\times 2$ positive definite matrix $\bm Q$ and vector $\bm t$ given by
\begin{equation}
\bm Q
=\begin{bmatrix}
Q_1 &Q\\
Q &Q_2
\end{bmatrix}
=\begin{bmatrix}
\bm x_1^T\bm x_1 & \bm x_1^T\bm x_2\\
\bm x_1^T\bm x_2 & \bm x_2^T\bm x_2
\end{bmatrix},
\qquad\text{and}\qquad
\bm t=\begin{bmatrix}
\bm x_1^T\bm h\\
\bm x_2^T\bm h
\end{bmatrix}.
\end{equation}
Let us call the integrand $\mathcal{F}$ and write
\begin{equation}
\mathcal{D}(\lambda,\varepsilon,\bm h,p)\propto\int_{\R^{2N}}d\bm x_1d\bm x_2\mathcal{F}(\bm Q,\bm t)
\end{equation}
Stating this form the right hand side is proportional to some (known) constant to (see~\cite{Fyodorov2019})
\begin{equation}\label{recast}
\int_{\R^2}\int_{\bm Q\succ 0}\mathcal{F}(\bm Q+\bm t  \bm t^T, h\bm t)(\det \bm Q)^{\frac{N-4}{2}}d\bm Qd \bm t
\end{equation}
with $\sqrt{\bm h^T \bm h}=h$, where the $\bm Q$ integral is over all $2\times 2$ positive definite symmetric matrices. After some additional manipulations the integrand in (\ref{recast}) is seen to be proportional to
\begin{align}
(Q_1Q_2-Q^2)^{\frac{N-4}{2}}\Big(&-h^2 \lambda ^2 \epsilon(h^2 \epsilon +2 \lambda) \Gamma(N-2,\lambda ^2)(Q^2-Q_1Q_2)-\Gamma(N,\lambda ^2)(2 i \lambda(i h^2 \epsilon +Q+t_1t_2) \nonumber\\
&+(t_1t_2+i h^2 \epsilon)^2+Q_1(Q_2+t_2^2)+Q_2 t_1^2)-\Gamma(n-1,\lambda ^2)(h^4\epsilon ^2 (Q_1Q_2+2 i \lambda  Q) \nonumber \\
&+2 h^2 \lambda  \epsilon  (2 i \lambda  Q+Q t_1 t_2+Q_1Q_2)
+\lambda ^2 (Q^2+2 Qt_1t_2-Q_2(Q_1+t_1^2)-Q_1t_2^2))+\Gamma(n+1,\lambda ^2)\Big) \nonumber \\
\times\exp\Big[\lambda^2+\frac{1}{2}(-&2 it_1t_2(h^2\epsilon +\lambda)-\frac{1}{2}t_1^2t_2^2-\sqrt{2p}(Q_1+Q_2+t_1^2+t_2^2)-2i \lambda Q-Q_1(Q_2+t_2^2)-Q_2t_1^2)\Big]
\end{align}
Before integrating out $\bm Q$ and $\bm t$ we firstly write the bi-quadratic term in the exponent with the Hubbard-Stratonovich transformation: $e^{-{t_1^2t_2^2}/{2}}\propto\int dye^{-y^2/2-iyt_1t_2}$. This additional step allows us to write the integrals over $t_2$ and $t_1$ as derivatives of one dimensional Gaussian integrals. Furthermore, the most convenient parametrization for the positive definite matrix $\bm Q$ is given by
\begin{equation}
\begin{bmatrix}
Q_1 & Q\\
Q & \frac{r^2+Q^2}{Q_1}
\end{bmatrix}
\end{equation}
where $r=\det^{1/2}(\bm Q)$ and measure $d\bm Q=2\frac{dQ_1}{Q_1}r dr dQ$ with $r>0,Q_1>0,Q\in\R$. After rescaling $Q_1\rightarrow\sqrt{2p}Q_1$ and $t_{1,2}\rightarrow (2p)^{1/4}t_{1,2}$ and integrating out $r,Q,t_2$ and $y$ we are left with
\begin{multline}\label{Cc}
\mathcal{D}(\lambda,\varepsilon,\bm h,p)\propto2^{\frac{N-1}{2}}\pi^{\frac{3}{2}}\int_{\R^{+}}dQ_1\int_{\R}dt_1\exp\Big[\frac{-h^4\epsilon ^2t_1^2-2 h^2\epsilon \lambda t_1^2  +\lambda ^2(Q_1+t_1^2+2)}{2(Q_1+t_1^2+1)}-p(Q_1+t_1^2)\Big]\frac{1}{(1+Q_1)^2}\\
\times Q_1^{\frac{N-3}{2}}(Q_1+t_1^2+1)^{-\frac{N}{2}-4}(b_0(Q_1+t_1^2+1)^2-b_1t_1^2(Q_1+t_1^2+1)(h^2 \epsilon +\lambda)+b_2(t_1^4(h^2 \epsilon +\lambda)^2+Q_1^2+Q_1(t_1^2+2)+t_1^2+1))
\end{multline}
where
\begin{align*}
s&=1+Q_1+t_1^2\\
u&=\lambda ^2 Q_1 \Gamma (N-1,\lambda ^2)-(Q_1+t_1^2) \Gamma(N,\lambda ^2)\\
v&=\lambda ^2 Q_1 \Gamma (N-1,\lambda ^2)-s\Gamma(N,\lambda ^2)\\
b_2&=-t_1^2s^2 \Gamma\left(\frac{N}{2}-1\right)(\lambda ^2 Q_1\Gamma(N-1,\lambda^2)-(Q_1+t_1^2) \Gamma(N,\lambda ^2))\\
b_1&=-2 t_1^2s \Gamma\left(\frac{N}{2}-1\right)(h^2 \epsilon +\lambda)(\lambda ^2Q_1 t_1^2\Gamma(N-1,\lambda ^2)-(t_1^2-1) s \Gamma(N,\lambda ^2))\\
b_0&=-h^4s^2t_ 1^2 u \epsilon ^2 \Gamma\left(\frac{N}{2}-1\right)+2 h^2s ( Q_ 1+1) t_1^2 v \epsilon  \Gamma\left(\frac{N}{2}-1\right)(h^2 \epsilon +\lambda)+2 \lambda  ( Q_1+1) s  t_1^2 v \Gamma\left(\frac{N}{2}-1\right)(h^2 \epsilon +\lambda)\\
&-2 h^2 \lambda s^2  t_1^2 u \epsilon  \Gamma\left(\frac{N}{2}-1\right)+( Q_ 1+1)^2 \bigg(\Gamma\left(\frac{N}{2}-1\right)(\Gamma(N,\lambda ^2)(s^2(h^4 s \epsilon ^2+N s- Q_ 1- t_1^2)+2 h^2 \lambda  s^2 \epsilon -\lambda ^2  Q_ 1 (s+1))\\
&-\Gamma(N-1,\lambda^2)(h^4  Q_ 1 \epsilon ^2(\lambda ^2 (s+1)+s)+2 h^2 \lambda   Q_ 1 \epsilon(\lambda ^2 (s+1)+s)-\lambda^2t_1^2(s-\lambda ^2  {Q_1}))+e^{-\lambda ^2} s^2 \lambda ^{2 N})\\
&+2 s \Gamma\left(\frac{N}{2}\right)(h^2 \lambda ^2Q_ 1 \epsilon(h^2 \epsilon +2 \lambda) \Gamma(N-2,\lambda ^2)+\Gamma(N-1,\lambda ^2)(-h^4{Q_1} \epsilon^2-2 h^2 \lambda   Q_ 1 \epsilon +\lambda ^2( Q_ 1+ t_1^2))\\
&-( Q_ 1+ t_1^2) \Gamma(N,\lambda^2)) \bigg)+ Q_ 1 s^2 u \Gamma\left(\frac{N}{2}-1\right)-\lambda ^2 s^2  t_1^2 u \Gamma\left(\frac{N}{2}-1\right)+s^2 u \Gamma\left(\frac{N}{2}-1\right).
\end{align*}
In the equalities above we introduced the relation $\Gamma(N+1,\lambda^2)=e^{-\lambda^2}\lambda^{2N}+N\Gamma(N,\lambda^2)$ and we got rid of the formally divergent terms by noticing $\Gamma(N+1,\lambda^2)-(N+\lambda^2)\Gamma(N,\lambda^2)+\lambda^2(N-1)\Gamma(N-1,\lambda^2)=0$. For the purpose of this work, in order to obtain the large $N$ behaviour of $\mathcal{D}(\lambda,\varepsilon,\bm h,p)$, it's more convenient to obtain the asymptotics from (\ref{Cc}) and to not proceed with the remaining integrations. Before doing that we recover the constant of proportionality as follows. First we notice that
\begin{equation}
\lim_{p\rightarrow+\infty}(2p)^{N/2}\mathcal{D}(\lambda,\varepsilon,\bm h,p)
=\E_\text{Gin}[\det(\lambda\bm 1_N+\varepsilon\bm h  \bm h^T-\bm \varXi)^2]
\end{equation}
Following the steps above, the latter expectation can be easily computed, this time keeping track of the constants of proportionality,
\begin{equation}
\E_\text{Gin}[\det(\lambda\bm 1_N+\varepsilon\bm h  \bm h^T-\bm \varXi)^2]=\lambda^{2N}+(N+h^4\varepsilon^2+2\lambda h^2\varepsilon)e^{\lambda^2}\Gamma(N,\lambda^2)
\end{equation}
In (\ref{Cc}), as $p\rightarrow+\infty$, the most relevant contribution to the double integral comes from $Q_1\rightarrow 0^{+}$ and $t_1\rightarrow 0$. In these limits, the second line of (\ref{Cc}) becomes equal to $\E_{Gin}[\det^2(\lambda\bm 1_N+\varepsilon\bm h\bm h^T-\bm \varXi)]e^{-\lambda^2}\Gamma(\frac{N}{2}-1)$. Therefore, for $p\rightarrow+\infty$, we must have
\begin{equation}
C_N\Gamma\Big(\frac{N}{2}-1\Big)2^{\frac{N-1}{2}}\pi^{\frac{3}{2}}\lim_{p\rightarrow+\infty}(\sqrt{2p})^N\int_{\R} dt_1\int_{\R^{+}}dQ_1\exp\Big[-p(Q_1+t_1^2)  -\frac{1}{2}(\varepsilon^2h^4+2\varepsilon\lambda h^2)t_1^2\Big]Q_1^{\frac{N-3}{2}}=1
\end{equation}
from which we obtain $C_N=(4\sqrt{2}\pi^{\frac{5}{2}}\Gamma(N-2))^{-1}$. This last step allows us to finally obtain, imposing $p\rightarrow0^{+}$, the desired expectation
\begin{align}
\E_{Gin}[|\det(\lambda\bm 1_N+\epsilon\bm h\bm h^T-\bm \varXi)|]
=&{}\frac{2^{-\frac{N}{2}-1}e^{-\lambda ^2}}{\sqrt{\pi}\Gamma (\frac{N+1}{2})}\int_{\R} dt_1\int_{\R^{+}}dQ_1
\exp\Big[\frac{-h^4\epsilon ^2t_1^2-2 h^2\epsilon \lambda t_1^2  +\lambda ^2(Q_1+t_1^2+2)}{2(Q_1+t_1^2+1)}\Big] \nonumber \\
\times Q_1^{\frac{N-3}{2}}(Q_1+t_1^2+1)^{-\frac{N}{2}-2}
\Big(&+\lambda ^{2 N}(h^4\epsilon ^2Q_1+(N-1)(Q_1+t_1^2+1))+2 h^2\epsilon\lambda ^{2N+1}Q_1 \nonumber \\
&+e^{\lambda ^2} \Gamma(N,\lambda ^2)(\lambda ^2((1-N)(Q_1+t_1^2)-h^4Q_1 \epsilon ^2)+(N-1)(h^4 \epsilon ^2(Q_1+1) \nonumber \\
&+N(Q_1+t_1^2+1))+2 h^2\epsilon \lambda (N-1) (Q_1+1) -2 h^2 \lambda ^3Q_1\epsilon)\Big) \label{EGin}
\end{align}
The result above is valid for any finite $N$,$\lambda,\varepsilon\in \C$ and $\bm h\in\R^N$.

\section{Asymptotic for the matrix average}

In this section, we address the large-$N$ asymptotic behaviour of (\ref{EGin}). Firstly, the global spectral distribution of the eigenvalues of $\bm \varXi/\sqrt N$, converges in probability to the circular law, i.e. to the uniform distribution over the unitary disc as $N\rightarrow+\infty$. Therefore, in order to investigate the different regimes arising for $N\to \infty$, from now on we make explicit the dependency of $N$ in $\lambda$ and $\varepsilon$, i.e. $\lambda\rightarrow\sqrt{N}\lambda$ and $\varepsilon\rightarrow\sqrt{N}\varepsilon$, and set $h=\abs{\bm h}=1$. Note that this scaling is already incorporated in the parameter choice~\eqref{parameter-choice}.

It be will clear from what follows that $\lambda$ is the main parameter since different regimes arise in correspondence to whether the value $|\lambda|$ is greater or less than unity. Indeed, $\abs{\lambda}=1$ corresponds to an evaluation at the edge discontinuity for the global spectral density. A nonzero $\varepsilon$ results in a rank-$1$ perturbation to $\bm \varXi$, which may result in a spectral outlier but it will not alter the location of the edge of global spectral density. From (\ref{EGin}) with the aforementioned parameter choice, we see that it is usefull to introduce
\begin{equation}
\mathcal{L}(q,t)=\frac{-\epsilon ^2t^2-2 \lambda  t^2 \epsilon +\lambda ^2+\lambda ^2(q+t^2+1)}{2 (q+t^2+1)}-\frac{1}{2} \log (q+t^2+1)+\frac{1}{2}\log (q)
\end{equation}
which collects the exponential terms and the first two term in the second line with power $N$ of (\ref{EGin}). The remaining terms in (\ref{EGin}) are collected by:
\begin{multline}
g(q,t)=q^{-\frac{3}{2}}(q+t^2+1)^{-2} (\lambda^{2N}N^{N}(N q \epsilon ^2+(N-1)(q+t^2+1))+2\lambda ^{2N+1} N^{N+1}q\epsilon\\
+e^{\lambda ^2 N} \Gamma (N,N \lambda ^2) (\lambda ^2 N ((1-N)(q+t^2)- N q \epsilon ^2)+(N-1)(N (q+1) \epsilon ^2+N (q+t^2+1))-2 \lambda ^3N^2 q \epsilon +2\lambda N (N-1)(q+1) \epsilon)
\end{multline}
For convenience, in $\mathcal{L}$ and $g$, we replaced $Q_1$ and $t_1$ with $q$ and $t$ respectively. As $N\to1$, the main contributions come from the set of saddle points of $\mathcal{L}(q,t)$. We see that the only feasible solution to $\nabla\mathcal{L}(q,t)=\bm 0$ is given by $q=\frac{1}{\lambda^2-1}$ and $t=0$ as it does not lead to $1+q+t^2=0$. At this point the Hessian of $\mathcal{L}(q,t)$, say $\bm H(q,t)$, is diagonal
\[
\bm H\Big(\frac{1}{\lambda^2-1},0\Big)=\begin{bmatrix}
-\frac{(\lambda^2-1)^4}{2\lambda^4} &0\\
0 &-\frac{(\varepsilon+\lambda)^2(\lambda^2-1)}{\lambda^2}
\end{bmatrix}
\]
and $\mathcal{L}(\frac{1}{\lambda^2-1},0)=-\frac{1}{2}+\lambda^2-\frac{1}{2}\log\lambda^2$. Thus, the Hessian $\bm H$ is negative definite only for $|\lambda|>1$. As mentioned above $\varepsilon$ does not play any role. We get two regimes

\begin{itemize}
\item We firstly assume that $|\lambda|>1$ and we observe that the point $(\frac{1}{\lambda^2-1},0)$ is contained in the domain of integration of (\ref{EGin}). Therefore, for $N\gg 1$, its neighbourhood gives the main contribution to the integral. Using the local approximations to $\mathcal{L}(q,t)$ and $g(q,t)$, i.e.
\begin{equation}
 \mathcal{L}(q,t)\approx\mathcal{L}\Big(\frac{1}{\lambda^2-1},0\Big)+\frac{1}{2}
\begin{bmatrix}
q-\frac{1}{\lambda^2-1}\\
t
\end{bmatrix}^T
 \bm H\Big(\frac{1}{\lambda^2-1},0\Big)
\begin{bmatrix}
q-\frac{1}{\lambda^2-1}\\
t
\end{bmatrix},
\quad\text{and}\quad
g(q,t)\approx g\Big(\frac{1}{\lambda^2-1},0\Big)
\end{equation}
We can integrate analitically $q$ and $t$ in (\ref{EGin}) as Gaussian integrals (see (\ref{GausId})). To leading order our Laplace approximation yields
\begin{equation}
\E_\text{Gin}[|\det(\lambda\bm 1_N+\epsilon\bm h\bm h^T-\bm \varXi)|]\sim N^{\frac{N}{2}}|\lambda|^{N-1}|\varepsilon+\lambda|(1+o(1)).
\end{equation}

\item For $|\lambda|<1$, we now observe that $(\frac{1}{\lambda^2-1},0)=\arg\max\mathcal{L}(q,t)\notin\R^{+}\times\R$. In order to extract the asymptotic for $N\gg 1$ we replace $\R^{+}\times\R$ with the subset $U(R)
\subset \R^{+}\times\R $ contained within the semicircle of radius $R$ and center $(0,0)$ in the first and fourth quadrants and the line connecting $(0,R)$ to $(0,-R)$, oriented counterclockwise. The unbounded domain is simply obtained imposing $R\rightarrow+\infty$. Since $(\frac{1}{\lambda^2-1},0)$ is not contained in $U(R)$ then $\mathcal{L}(q,t)$ necessarily reaches its maximum in correspondence of an accumulation point of $U(R)$. From the structure of $\mathcal{L}$, and by parametrizing the semicircle, it turns out that such point is $(R,0)$ with $\mathcal{L}(R,0)=1/2((R+2)\lambda^2/(1+R)+\log(R/(R+1)))$. The boundary of $U(R)$, i.e. $\partial U(R)$, is smooth and differentiable with curvature $R^{-1}$ in $(R,0)$ where the outward normal vector is simply $\bm n=(1,0)$. Therefore by making use of the divergence theorem we have (see~\cite{BH1986})
\begin{equation}\label{DS}
\E_{Gin}[|\det(\lambda\bm 1_N+\epsilon\bm h\bm h^T-\bm \varXi)|]=C_{N,\lambda}\lim_{R\rightarrow+\infty}\oint_{\partial U(R)}\frac{d\ell e^{N\mathcal{L}(\ell)}}{N} \frac{g(\ell)}{|\nabla\mathcal{L}|^2}\nabla\mathcal{L}(\ell)\cdot\bm n(\ell)+O(e^{N\mathcal{L}(R,0)}N^{-2})
\end{equation}
where $C_{N,\lambda}=\frac{2^{-\frac{N}{2}-1}e^{-\lambda ^2}}{\pi^{\frac{1}{2}}\Gamma (\frac{N+1}{2})}$. Expanding the terms above around the point $(R,0)$ yields
\begin{equation}\label{Gap}
\E_\text{Gin}[|\det(\lambda\bm 1_N+\epsilon\bm h\bm h^T-\bm \varXi)|]\approx C_{N,\lambda}\lim_{R\rightarrow+\infty}\sqrt{\frac{2\pi}{N^3}}e^{N\mathcal{L}(R,0)}g(R,0)\big(\sqrt{\big|f(R,0)+R^{-1}w(R,0))\big|}\big)^{-1}
\end{equation}
where: $w(q,t)=\big|\nabla\mathcal{L}(q,t)\big|^3$, $f(q,t)=\frac{\partial^2\mathcal{L}(q,t)}{\partial q\partial t}(\frac{\partial\mathcal{L}(q,t)}{\partial t})^2-2\frac{\partial^2\mathcal{L}(q,t)}{\partial q\partial t}(\frac{\partial\mathcal{L}(q,t)}{\partial t}\frac{\partial\mathcal{L}(q,t)}{\partial q})+\frac{\partial^2\mathcal{L}(q,t)}{\partial q\partial t}(\frac{\partial\mathcal{L}(q,t)}{\partial q})^2$. We observe that $f$ is the leading term in the square root of (\ref{Gap}) as $f(R,0)=O(R^{-5})$ while $w(R,0)=O(R^{-6})$. By saddle point approximation, one also gets $\Gamma(N,N\lambda^2)\approx N^N e^{-N}\int_{\lambda^2}^{+\infty}due^{-\frac{N}{2}(u-1)^2}$ and $\erfc(\frac{\sqrt{N}(\lambda^2-1)}{\sqrt{2}})\approx 2$ for large $N$ and $|\lambda|<1$. After removing the remaining subleading terms in $N$ from (\ref{Gap}) we obtain
\begin{equation}
\E_{Gin}[|\det(\lambda\bm 1_N+\epsilon\bm h\bm h^T-\bm \varXi)|]
\sim \sqrt{2} N^{\frac{N}{2}} e^{\frac{N}{2} (\lambda ^2-1) } \sqrt{\epsilon ^2+2 \lambda  \epsilon +1}(1+o(1)).
\end{equation}

\item For completeness, we now consider the edge of the spectrum, $|\lambda|=1$. We further rescale $\lambda\rightarrow 1+\frac{\lambda}{\sqrt{N}}$ and $\epsilon\rightarrow 1+\frac{\epsilon}{\sqrt{N}}$. We assume that the new variables $\lambda$ and $\epsilon$ are of order one. Following the steps above and eq(\ref{DS}) with $N\gg 1$, by performing a saddle point approximation
we obtain:
\begin{itemize}
\item for $\lambda<0$:
\begin{equation}
\E_{Gin}[|\det(\lambda\bm 1_N+\epsilon\bm h\bm h^T-\bm\varXi)|]\sim \sqrt{2}N^{\frac{N}{2}}e^{\sqrt{N}\lambda+\frac{\lambda^2}{2}}\erfc(\sqrt{2}\lambda)(1+o(1))
\end{equation}
\item for $\lambda>0$:
\begin{equation}
\E_{Gin}[|\det(\lambda\bm 1_N+\epsilon\bm h\bm h^T-\bm\varXi)|]\sim 2\sqrt{N}(\sqrt{N}+\lambda)^{N-1}(1+o(1))
\end{equation}
\end{itemize}

\end{itemize}

\section{Mean Number of fixed Points}
We can now substitute the formulas above for $\E_{Gin}[|\det(\lambda\bm 1_N+\epsilon\bm h\bm h^T-\bm\varXi)|]$ in the definition of the density $\widehat\rho_{\mu}(r>0)$ replacing $\lambda$ and $\varepsilon$ with $\frac{\mu}{\sqrt{C'(r^2)}}$ and $\mu\sqrt{\frac{C(r^2)}{\Delta(r^2)}}(1-\frac{C' (r^2)}{C(r^2)}r^2)$ respectively. From the monotonicity of $C(r^2)$ the edges regimes do not play any role as they correspond to null measure sets in $\R^N$. Therefore, in the large $N$ limit, $\E_{Gin}[|\det(\bm\varXi-\mu\sqrt{N}\bm D)|]$ is given by $N^{N/2}|\frac{\mu}{\sqrt{C'(r^2)}}|^{N-1}|\mu\sqrt{\frac{C(r^2)}{\Delta(r^2)}}(1-\frac{C'(r^2)}{C(r^2)}r^2)|$ and by $\sqrt{2}N^{N/2}e^{\frac{N}{2}\big(\frac{\mu^2}{C'(r^2)}-1\big)}\sqrt{\mu^2\frac{C(r^2)}{\Delta(r^2)}\big(1-\frac{C'(r^2)}{C(r^2)}r^2\big)^2-\frac{\mu^2}{C'(r^2)}+1}$ for $|\mu/\sqrt{C'(r^2)}|>1$ and $|\mu/\sqrt{C'(r^2)}|<1$ respectively.
\begin{itemize}
\item We start considering $\mu<\sigma_1$. In this case:
\begin{equation}\label{case1}
\widehat\rho_{\mu}(r)=
\frac{\sqrt{2N}}{\sqrt{\pi}}\frac{e^{\frac{N}{2} L_{II}(r)}}{r} \frac{\Delta^{1/2}(r^2)}{\sqrt{C(r^2)C'(r^2)}}\sqrt{\mu^2\frac{C(r^2)}{\Delta(r^2)}\Big(1-\frac{C'(r^2)}{C(r^2)}r^2\Big)^2-\frac{\mu^2}{C'(r^2)}+1}\mathbbm{1}(0<r<R)
\end{equation}
where $L_{II}(r)=\log r^2-\frac{\mu^2 r^2}{C(r^2)}+\log\frac{C'(r^2)}{C(r^2)}+\frac{\mu^2}{C'(r^2)}$
and $\mathbbm{1}(x)$ is the indicator function with value $1$ iff the condition $x$ is satisfied.

\item Let's assume now $\mu>\sigma_1$. This requires to investigate separately the contributions of the two asymptotic expressions for $\E_{Gin}[|\det(\bm\varXi-\mu\sqrt{N}\bm D)|]$. Let $r_{-}$ be the solution of $C'(r^2)=\mu^2$ and let's introduce $L_I(r)=+1+\log\mu^2+\log r^2-\frac{\mu^2r^2}{C(r^2)}-\log C(r^2)$, for $N\gg 1$, we have:
\begin{multline}
\label{case2}
\widehat\rho_{\mu}(r)=\frac{\sqrt{N}}{\sqrt{\pi}}\frac{e^{\frac{N}{2} L_I(r)}}{r}\Big|1-\frac{C'(r^2)}{C(r^2)}r^2\Big|\mathbbm{1}(0<r<r_{-})+\\
+\frac{\sqrt{2N}}{\sqrt{\pi}}\frac{e^{\frac{N}{2} L_{II}(r)}}{r} \frac{\Delta^{1/2}(r^2)}{\sqrt{C(r^2)C'(r^2)}}\left(\sqrt{\mu^2\frac{C(r^2)}{\Delta(r^2)}\Big(1-\frac{C'(r^2)}{C(r^2)}r^2\Big)^2-\frac{\mu^2}{C'(r^2)}+1}\right)\mathbbm{1}(r_{-}<r<R)
\end{multline}
For $r>r_{-}$ and sufficiently large $N$, the second term in (\ref{case2}) prevails over the first one as $\frac{\mu^2}{C'(r^2)}-\log\frac{\mu^2}{C'(r^2)}-1>0$.
\end{itemize}

\section{Plots and numerics}

This section contains plots and numerical simulations intended to illustrate and supplement the main conclusions of the letter.

\begin{figure}[ht]
\includegraphics[width=.9\textwidth]{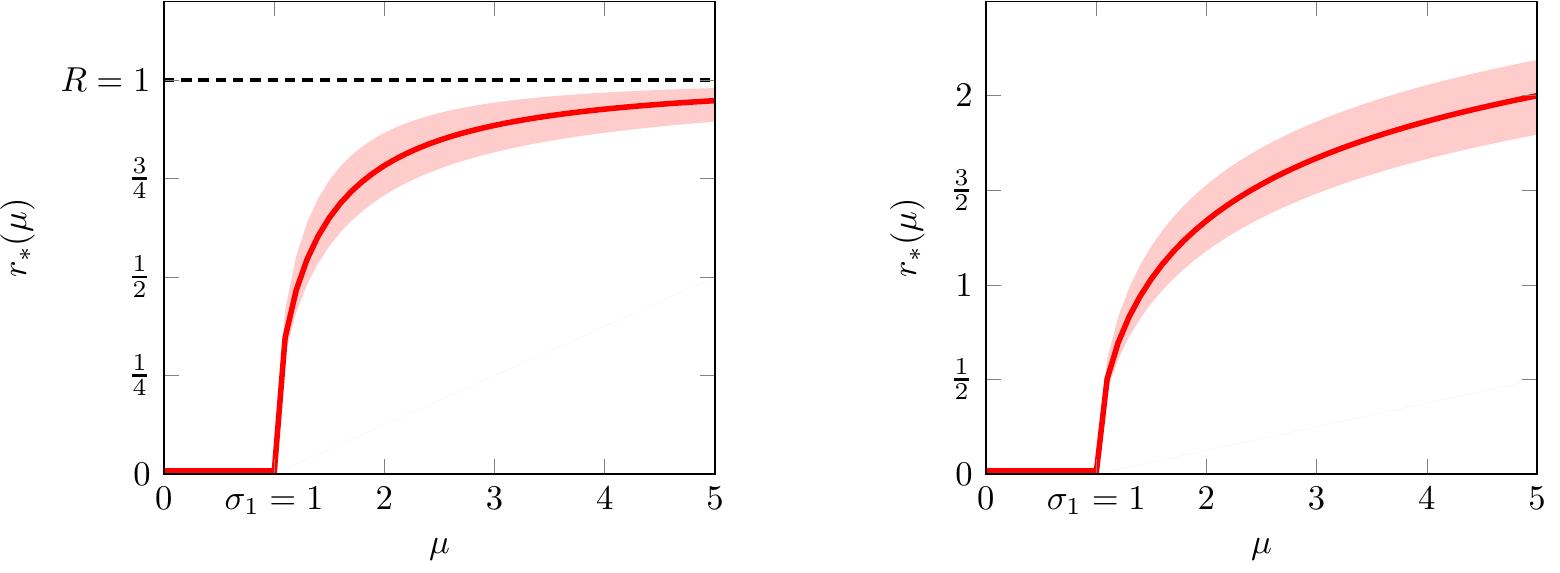}
\caption{}\label{r}
\end{figure}

Figure~\ref{r} shows the radius $r_*(\mu)$ as function of the parameter $\mu$ for a correlation $C(r^2)=r^2/(1-r^2)=r^2+r^4+r^5+\cdots$ (left panel) and $C(r^2)=e^{r^2}-1=r^2/1!+r^4/2!+r^6/3!+\cdots$ (right panel). The former has radius of convergence $R=1$ while the latter has infinite radius of convergence $R=\infty$; both has $\sigma_1=1$. The dark red curve shows $r_*(\mu)$ while the light red region shows $r\in(r_-(\mu),r_+(\mu))$. Below the threshold, i.e $\sigma_1<\mu$, we have $r_*(\mu)=r_\pm(\mu)=0$. Beyond the threshold, i.e $\sigma_1>\mu$, we know that $r_\pm(\mu)$ are strictly monotonically increasing function and that $r_*(\mu),r_\pm(\mu)$ tends to $R$ as $\mu\to\infty$.

\begin{figure}[ht]
\includegraphics[width=.9\textwidth]{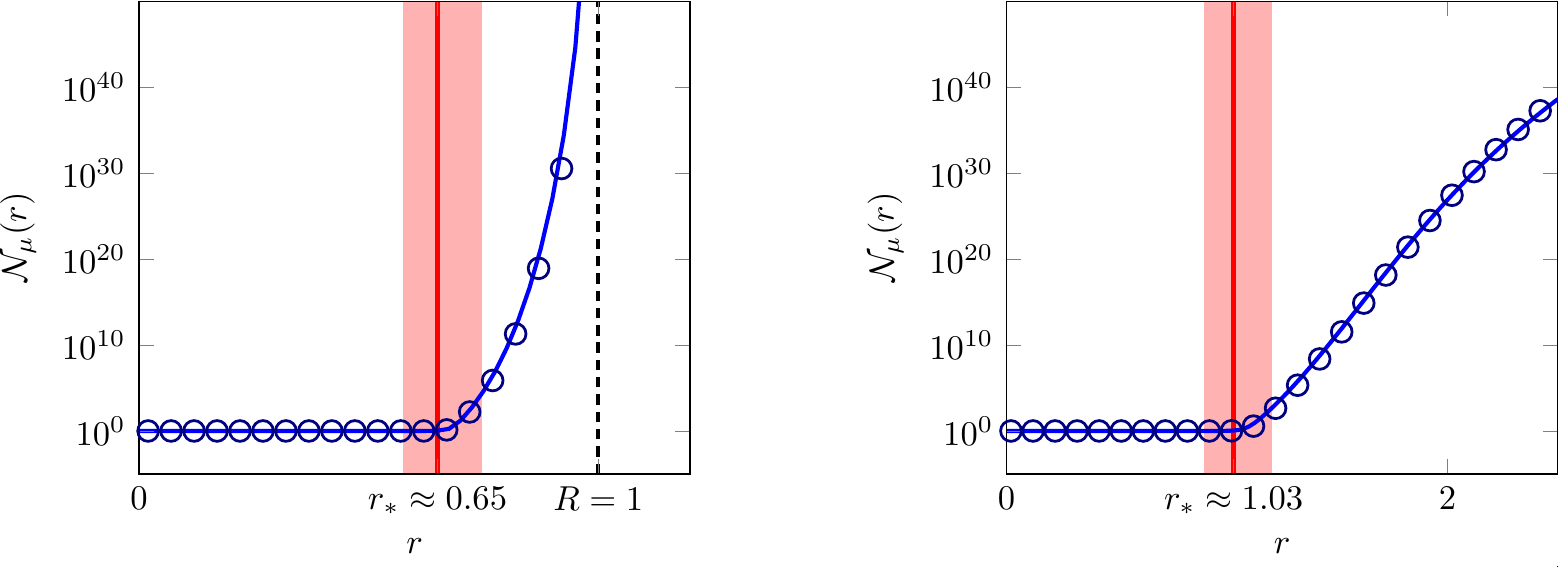}
\caption{}\label{det-num}
\end{figure}

Figure~\ref{det-num} shows the mean number of fixed points within a ball of radius $r$ centred at the origin, $\mathcal N_\mu(r)$, with $N=100$. Like figure~\ref{r}, the left plot is with $C(r^2)=r^2/(1-r^2)$, $r<R=1$, and the right plot is with $C(r^2)=e^{r^2}-1$, $r<R=\infty$. Both plots have $\sigma_1=1$ and $\mu=3/2$ and therefore represent the phase in which the origin is locally stable. The dark red curve shows $r_*(\mu)$ while the light red region shows $r\in(r_-(\mu),r_+(\mu))$. Consistent with figure~\ref{r}, the critical radii are $r_*(\mu=3/2)\approx0.65$ and $r_*(\mu=3/2)\approx1.03$ for the left and right plots, respectively. The blue curves shows the mean number of fixed points within a ball of radius $r$ using the asymptotic formulae for the spherical density~\eqref{rho-asymp-} and~\eqref{rho-asymp+}, while the blue data points are found using the finite-$N$ formula~\eqref{kac-rice-full} with the matrix-average evaluated numerically. It is seen that the numerical data and the asymptotic formulae are in complete agreement. Within any ball of radius $r<r_*$ the origin is the only fixed point, but for $r>r_*$ the mean number of fixed points quickly grows many orders of magnitudes (for larger $N$ this growth becomes even steeper). In the two cases illustrated in figure~\eqref{det-num}, $\mathcal N_\mu(r)$ tends to infinity as $r\to R$ even for finite $N$, i.e. the total number of fixed points is infinite. However, this is not always the case, e.g. consider a correlation function described by a terminating series $C(r^2)=\sigma_1^2r^2+\cdots+\sigma_M^2r^{2M}$. In this case, the total number of fixed points is bounded from above by $M^N$ with probability one (note that this upper bound grows exponentially with $N$).

\begin{figure}[ht]
\includegraphics[width=.9\textwidth]{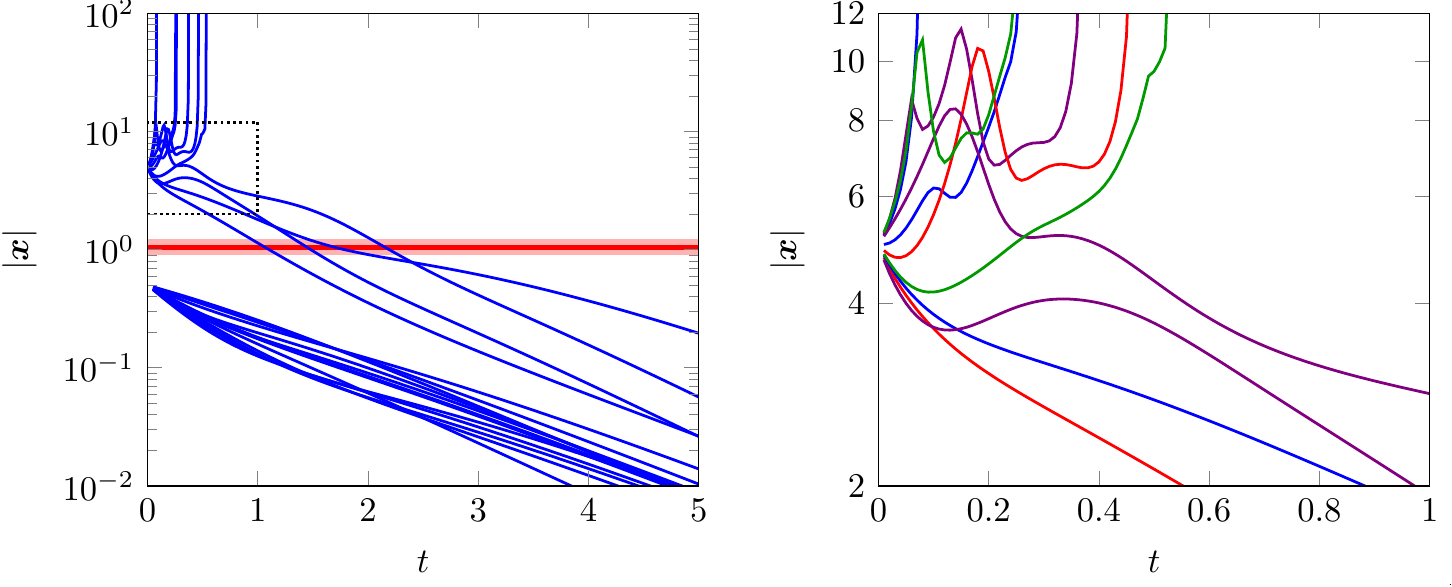}
\caption{}\label{paths}
\end{figure}

Figure~\ref{paths} shows the Euclidean distance to the origin for different paths with initial positions not at the origin with $\sigma_1=1<3/2=\mu$, i.e. the scenario in which the origin is locally stable. Before we give the explicit values which we have used for the plots in figure~\ref{paths}, we will briefly describe the general procedure.
First, we have chosen one realisation of the random vector field as $\phi_n(\bm x)=\sum_{k=1}^M\frac1{\sqrt{k!}}\sum_{i_1,\ldots,i_k}\xi_{n,i_1,\ldots,i_k}x_{i_1}\cdots x_{i_k}$ with $n=1,\ldots,N$ and $\xi_\bullet$ i.i.d. standard Gaussian random variables. We emphasise that we use the same realisation of $\bm\phi$ for all paths. Second, we have chosen $p$ points uniform at random on the unit $(N-1)$-sphere; let us denote the corresponding unit vectors by $\bm e_1,\ldots, \bm e_p$. Initial conditions for our $p$ paths are chosen as $\bm x^1(0)=\epsilon\bm e_1,\ldots,\bm x^p(0)=\epsilon\bm e_p$ for some positive constant $\epsilon$, which we may interpret as the radial component of a perturbation at time $t=0$. By numerically evolving the system according to the dynamical law $d\bm x/dt=-\mu\bm x+\bm\phi(\bm x)$, we can then investigate the dependence of paths on the radial component of initial perturbation by varying the constant $\epsilon$. For $\epsilon\ll r_*$, all paths are expected to decay back to the origin independent of the direction of the perturbation, while $\epsilon\gg r_*$ the behaviour of the paths is expected to be sensitive to the direction. For the plots shown on~\ref{paths} we have $N=M=4$ and $p=10$. On the left plot, the blue curves shows the Euclidean norm $\abs{\bm x}$ for $\epsilon=0.5$ and $\epsilon=5$; the unit vectors and the random landscape are the same for both values of epsilon. Like on figure~\ref{r} and~\ref{det-num}, the dark red line shows $r_*(\mu)$ while the light red region shows $r\in(r_-(\mu),r_+(\mu))$. The right plot on figure~\ref{paths} is an enhanced version of the dotted box indicated on the left plot. The colours of paths on the right plot are included to better distinguish between different paths and they has no further significance.  We see that for $\epsilon=1/2<r_*$ all paths decays back to the origin, but for $\epsilon=5>r_*$ some paths decays while other diverge. We also done numerical checks for different values of $\epsilon$, $\mu$, $N$, $M$ as well as for different realisations of the random field $\bm\phi$ (not shown on plots), which all show agreement with the hypothesis of resilience to perturbations $\epsilon\ll r_*$ but sensitivity to perturbations $\epsilon\gg r_*$. Increasing (decreasing) $\epsilon$ generally results in less (more) paths which decays back to the origin. We note that for the plots shown on figure~\ref{paths} we have $N=M=4$, hence we know that there are less than $M^N=64$ fixed points. Except for the origin, most fixed points are expected to be unstable, so if a path is sufficiently far from the origin they are expected to feel an effective repulsion and therefore diverge. Closer to the origin dynamics is expected to be more complicated as illustrated on the right plot on figure~\ref{paths}.

\end{document}